\documentclass[a4paper,11pt]{article}
\pdfoutput=1 

\usepackage{jheppub,graphicx} 

\usepackage[all]{xy}
\usepackage[T1]{fontenc} 
\usepackage{subfig}

\newcommand{\PP}{\mathbb{P}}
\newcommand{\CC}{\mathbb{C}}
\newcommand{\XX}{$\mathcal{X}$}
\newcommand{\A}{$\mathcal{A}$}
\DeclareMathOperator{\Conf}{Conf}
\DeclareMathOperator{\Gr}{Gr}

\def\beq{\begin{equation}}
\def\eeq{\end{equation}}
\def\beqa{\begin{eqnarray}}
\def\eeqa{\end{eqnarray}}
\newcommand{\vev}[1]{\langle #1\rangle}

\title{\boldmath Cluster Algebras in Kinematic Space of Scattering Amplitudes}

 \author{Marcus A. C. Torres}
 \affiliation{IMPA,\\Est. Dona Castorina 110,\\ 22460-320 Rio de Janeiro-RJ, Brazil}

\emailAdd{mtorres@impa.br}

\abstract{We clarify the natural cluster algebra of type A that exists in a residual and tropical form in the kinematical space as suggested in 1711.09102 by the use of triangulations, mutations and associahedron on the definition of scattering forms. We also show that this residual cluster algebra is preserved in a hypercube (diamond) necklace inside the associahedron where cluster sub-algebras $(A_1)^n$ exist.  This result goes in line with results with cluster poligarithms in 1401.6446 written in terms of $A_2$ and $A_3$ functions only and other works showing the primacy of $A$ cluster sub-algebras as data input for scattering amplitudes.}

\begin{document} 
\maketitle
\flushbottom

\section{Introduction}
\label{sec:intro}

In \cite{Golden:2013xva} cluster algebras made quite an appearance in the studies of Scattering Amplitudes. There, the authors showed that a judicious choice of kinematic variables was one of the main ingredients in a large simplification of the previously calculated two-loops six particle MHV remainder function $\mathcal{R}_n^{(2)}$  \cite{DelDuca:2009au,DelDuca:2010zg,Zhang:2010tr} of  $\mathcal{N}=4$ supersymmetric Yang-Mills (SYM). This choice is related to the cluster structure that is intrinsic to the kinematic configuration space $\Conf_n(\PP^3)$ of $n$ external particles. Its cluster structure selects the appropriate cross ratios (directly related to $\mathcal{X}$-coordinates in the cluster algebras) to be used in the motivic amplitudes. Intriguingly, some cluster algebras define arguments so suitably to some polylogarithmic functional equation as the famous Abel's pentagon dilogarithm identity and a recently found trilogarithm relation \cite{Golden:2013xva}, showing that the use of cluster coordinates as arguments in remainder functions may be the appropriate way to simplify the long polylogarithm expressions.
In agreement, many other studies with symbols used cluster coordinates as arguments to polylogarithm expressions \cite{Dixon:2011pw,Dixon:2011nj,CaronHuot:2011kk}.

The  cluster algebra in $\Conf_n(\PP^3)=\Gr(4,n)/(\CC^*)^{n-1}$ is the cluster algebra of its  Grassmannian $\Gr(4,n)$ \cite{scott2006grassmannians}. Since the dimension of the configuration space is $3n-15$, the  non-trivial cases are for $n\geq 6$.  $\Conf_6(\PP^3)$ has  cluster algebra  of type $A_3$ and at $n=7$ it has cluster algebra of type $E_6$, where the subscript numbers are the rank of the cluster algebra, i. e., the number of cluster variables in one cluster. The capital letters is how the quiver representation of the cluster algebra compares to Dynkin diagrams. This comparison works for all finite type cluster algebras and they go beyond relating root system and cluster variables \cite{Fomin:ky}.

The use of the cluster structure of  $\Conf_n(\PP^3)$ worked efficiently for $n=6, 7$, but for $n$ greater than 7, the cluster algebra is of infinite type and we cannot count all of its \A-coordinates, \XX-coordinates and clusters. 

Some questions were raised in  \cite{Golden:2013xva} and in subsequent work \cite{Golden:2014xqa} about a choice of a subset of cluster variables to be used in the motivic amplitude and the exclusive use of polylogarithm functions with entries on $A_n$ cluster algebras \cite{Golden:2014xqa,Parker:2015cia}. At $n=$ 6 and at $n=$  7 only $3/5$ of all cluster \XX-coordinates (or $3/10$ if  inverse \XX-coordinates are considered separately)  of their respective cluster algebra show up in the two-loop MHV motivic amplitude.


In \cite{Golden:2014xqa}, the authors built polylogarithm functions of weight 4 with entries on $A_2$ and $A_3$ cluster algebras. In \cite{Parker:2015cia}  bases of polylogarithm functions of any weight k were built using symbol alphabets in $A_n$ cluster algebra and they affirmed that similar basis cannot be built using other cluster algebras. Also in this paper, the authors explained the interesting ratio $3/5$ of the subset of  used \XX-coordinates in the symbol alphabet of the motivic amplitude out of the total of \XX-coordinates in the cluster algebra of $A_3$ and $E_6$ for $n=6,7$, respectively. It is simply because for $A_3$ cluster algebra only 9 out of 15 (or 30) \XX-coordinates are multiplicative independent. Since the $\mathcal{R}_7^{(2)}$ amplitude for n=7 can be built using $A_3$ cluster functions \cite{Golden:2014xqa,Golden:2014xqf}, we expect that this explains this same ratio at n=7. From this answer, we raise and intend to answer another question: can we find general bases of multiplicatively independent \XX-coordinates for all $A_n$ cluster algebra? A possible answer already appeared in \cite{Torres:2013vba}, as the set of \XX-coordinates of all snake clusters (up to inversion). That's what we expect due to Laurent phenomena in cluster algebras and the correspondence of the cluster variables and the $A_n$ root system. 
 
 Other interesting properties of cluster algebra in scattering amplitudes were hinted in \cite{Arkani-Hamed:2017mur}. There, the authors showed that a $n$-point planar tree level scattering amplitude of a  bi-adjoint $\phi^3$ theory can be recovered from scattering forms built on kinematic variables that represent diagonals of a n-polygon and that each triangulation of this polygon represents a Feynman diagram. 
 
 We show in section \ref{cluster kinematical} that a general $A_n$ cluster algebra is residually present in the kinematic space $\mathcal{K}_n$ spanned by planar kinematic variables $X_{ij}$ of a n-point amplitude in certain limit of high energy. It remains to be understood if this residual $A_n$ cluster algebra present in the kinematic space justify their primacy in building cluster functions for amplitudes. These are somewhat different cluster algebras in comparison to the ones existing  in the kinematic configuration space $\Conf_n(\PP^3)$ which is built from kinematic data in momentum twistor space. 

Cluster algebras appeared also in SYM scattering amplitudes at the integrand level \cite{ArkaniHamed:2012nw}, but we make no connection with it here. We discuss here cluster algebra as a unique property of the kinematic space of an amplitude and we do not discuss how it is preserved at loop level or at non MHV cases (Yang-Mills). The tree-level bi-adjoint $\phi^3$ amplitude in \cite{Arkani-Hamed:2017mur} is free from these obstacles and this may the reason why it makes evident so many cluster algebra features, to the point of its respective amplituhedron becomes an $A_n$ associahedron or Stasheff polytope.

In section \ref{ca} we review basic elements of cluster algebras and $A_n$ cluster algebras. In section \ref{cluster kinematical} we present the kinematic variables and the corresponding cluster algebra. In section \ref{Kn vs Confn} we compare the cluster algebras of $\mathcal{K}_m$ and $\Conf_n(\PP^3)$ using a standard basis in the literature \cite{Drummond:2007au,Anastasiou:2009kna} of cross-ratios  and and we show that they are directly related to coordinates of $A$ type cluster algebras in both cases. We finish this work in section \ref{conclusion} with a conclusion that raises more issues for future work, since many things  about cluster algebras in scattering amplitudes are not clear yet.

\section{Cluster Algebras}\label{ca}
The subject of cluster algebras was first presented in a accessible way to the physics community of scattering amplitudes in \cite{Golden:2013xva}. For further information,  the standard references in the area are \cite{Gekhtman:be,Keller:2008lt,Fomin:ky,Fomin:hy}. The reference for cluster algebras in Grassmannians is \cite{scott2006grassmannians}.  We  review here only basic concepts and  terms.

We are only interested in the finite type cluster algebras.  Such cluster algebras have a finite number $n$ of distinct  generators (cluster variables), that is grouped in a finite number of clusters (sets) of equal size $m<n$ and that relate to each other by exchange relations where one of the cluster variables is replaced by (mutates to)  another cluster variable outside the cluster. These exchange relations  can be codified within each cluster by associating them with oriented quivers. From the quiver associated to a cluster we can define \XX-coordinates related to each cluster variable in that cluster. 

A cluster may contain a subset of frozen variables (cluster coefficients)  that do not mutate and stays the same in all clusters. The number $m$ of  cluster variables (not frozen ones) in a cluster is the rank of the cluster algebra. We call both cluster variables and cluster coefficients as \A-coordinates. 

Quivers are built with arrows connecting vertices. Such vertices in a quiver are identified with \A-coordinates while arrows  define exchange relations among \A-coordinates and the \XX-coordinates of each vertex in the quiver.  Quivers are such that loops and two-cycles are not allowed. Loops are arrows that have same origin and target and two-cycles are a pair of arrows with opposite direction connecting the same two vertices. When a two-cycle appears after a mutation, the arrows ``cancel each other'' and disappear.

A mutation of a cluster variable in a cluster, or vertex $k$ in the corresponding quiver transforms it to a new quiver according to the following operations:
\begin{itemize}
\item for every pair of arrows $i\rightarrow k$ and $k\rightarrow j$, add a new arrow $i\rightarrow j$,
\item reverse all arrows that target k or depart from k,
\item proceed with all two-cycle cancellation.
\end{itemize}

A theorem \cite{Fomin:ky}  classifies all finite type cluster algebras according to simply  laced Lie Algebras. It states that given a  finite type cluster algebra, their clusters have quivers  that are mutation equivalent to a Dynkin diagram of a Lie Algebra, via identification of the principal part of its quiver. The principal part of a quiver is the quiver without frozen variables and arrows to or from them. 

There can be more than 1 arrow between 2 vertices and a number can be added on top of each arrow for cases of multiple arrows. An skew symmetric adjacency matrix $(b_{ij})$ can be defined from the quiver, where
\beq
b_{ij}= \# arrows (i\rightarrow j)-\#arrows (j\rightarrow i)
\eeq

A cluster variable in vertex $k$, $a_k$, mutates to $a'_k$ according to the exchange relation:
\beq
a'_k a_k= \prod_{i| b_{ki}>0}a_i^{b_{ki}}+\prod_{i| b_{ki}<0}a_i^{-b_{ki}}
\eeq 
 
 where $i$ goes through  cluster coefficients too. In a quiver, for every vertex  correspondent to a cluster variable we define its \XX-coordinate $x_i$ in terms of \A-coordinates $a_j$,
\beq
x_i= \prod_{j\neq i}a_j^{b_{ij}}
\label{xdef}
\eeq

We remark that when a mutation occurs in one vertex forming a new cluster, the adjacency matrix changes accordingly and following eq. (\ref{xdef}) the new cluster will have different \XX-coordinates. The \XX-coordinate of vertex $i$ under mutation, mutates from $x_i$ to $x_i^{-1}$. Throughout the paper, we will not count a \XX-coordinate  and its inverse as independent \XX-coordinates.

A useful construction associated to a finite type cluster algebra is a generalized associahedron. Such construction represents the cluster algebra as a polytope with clusters being represented by vertices and mutations between clusters being represented by edges connecting vertices. For a rank $r$ cluster algebra, each vertex is parametrized by the $r$ \XX-coordinates of the represented cluster and from each vertex departs $r$ edges. 

 In type $A$ cluster algebra  the generalized associahedron is called Stasheff polytope \cite{Fomin:ky}, which will be used here, since we will be always dealing with type A cluster algebras.

Naming the rank of a Stasheff polytope as the rank of the corresponding cluster algebra, an interesting property of a Stasheff polytope  is that its boundaries are made of lower rank Stasheff polytopes corresponding to local cluster subalgebras. 

Rank one $A_1$  and rank two  $A_1\times A_1$ and $A_2$ cluster algebras are associated to the smallest Stasheff polytope which are dimension one edge and dimension two quadrilateral and pentagonal faces, of an $A_3$ polytope, for example.

\subsection{\texorpdfstring{$A_n$}{An}  Cluster Algebras}
  
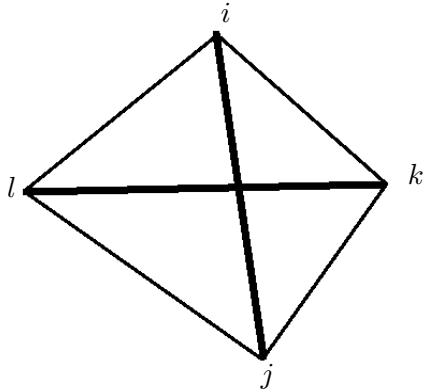
\begin{figure}
\begin{picture}(150,150)(-80,0)
\put(0,0){\includegraphics[scale=0.4]{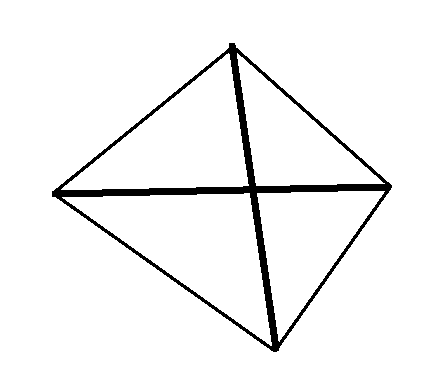}}
\put(95,136){$i$}
\put(165,75){$k$}
 \put(110,0){$j$}
 \put(15,70){$l$}
\end{picture}
\caption[diagonals]{ Diagonals $\overline {ij}$ and $\overline {kl}$ define two possible triangulations for the quadrilateral ikjl.}
\label{triangulation}
\end{figure}

This finite class of cluster algebras has some interesting features. First, it corresponds to the cluster algebra of $\Gr(2, n+3)$ where the \A-coordinates are their Pl\"ucker coordinates. Second, there is a geometric representation that replaces the use of quivers. The $A_n$ cluster algebra is 
 geometrically  represented  by triangulations of a polygon \cite{Fomin:ky,Keller:2008lt} in the following way:
\begin{itemize}
\item a cluster is associated to a  triangulation of a polygon ${\bf P}_{n+3}$ with $(n+3)$ edges, using $n$ diagonals such that no diagonal cross another one.
\item \A-coordinates $\vev{ij}$ (with $i < j$) of a cluster correspond to edges linking vertices i and j of the polygon. They are frozen variables when they correspond to side edges ($j=i+1$) and cluster variables when correspond to diagonals ($j>i+1$) of the polygon triangulation. There are $\frac{n(n+3)}{2}$ of them. Note that  $\vev{ij}=-\vev{ji}$, as Pl\"ucker coordinates and we avoid considering $\vev{ij}$ and $\vev{ji}$ separate \A-coordinates with the condition $i<j$.

\item  a mutation is associate to changing one diagonal to another diagonal such that both are diagonals of the same quadrilateral (figure \ref{triangulation}). 
\end{itemize}

The \A-coordinate $\vev{ij}$ in figure \ref{triangulation} mutates to the \A-coordinate $\vev{kl}$ according to the Pl\"ucker relation:
\begin{equation}\label{Plucker}
\vev{ij}\vev{kl}=\vev{ik}\vev{jl}+\vev{il}\vev{kj},
\end{equation}
for all $i<k<j<l\leq n+3$.

We can also identify the \XX-coordinates of a given cluster by the associated polygon triangulation.
For $i<k<j<l$, the diagonals of figure (\ref{triangulation})  represent two \A-coordinates $\vev{ij}$ and $\vev{kl}$, one is a  mutation from another. Their \XX-coordinates are given by the sides of the quadrilateral ikjl:
\beq
x_{ij}=x_{kl}^{-1}=\frac{\vev{ik}\vev{jl}}{\vev{kj}\vev{il}}
\label{ijcondition}
\eeq

   
   
 \subsection{Snake triangulations}
 The number of clusters in an $A_n$ cluster algebra is given by the number of  triangulations of the polygon ${\bf P}_{n+3}$ and it is given by the Catalan number $C_{n+1}=\frac{1}{n+1}\binom{2n+2}{n}$. While this number grows exponentially with n, the number of cluster variables $\tfrac{n(n+3)}{2}$ grows only quadratically. There is a much smaller set of $n$ clusters that contain together all cluster variables and they are  easily identifiable by their geometric realization as snake triangulations of ${\bf P}_{n+3}$ (zig-zag in \cite{scott2006grassmannians}).  They are the $n$ nodes of a hypercube (nicknamed diamond) necklace inside of the $A_n$ Stasheff polytope \cite{Torres:2013vba}. In fact, only half of the nodes (or clusters)  are necessary. $\lceil\tfrac{n+3}{2}\rceil$ consecutive nodes of the $A_n$ associahedron contain all cluster variables. $\lceil x\rceil$ represents the ceiling function that returns the smallest integer larger than or equal to x. 
 
\begin{figure}
\centering
\subfloat[$A_3$ necklace made with $A_1$ line and $A_1\times A_1$ quadrilateral beads.\label{fig:sub1}]{\includegraphics[width=0.52\linewidth]{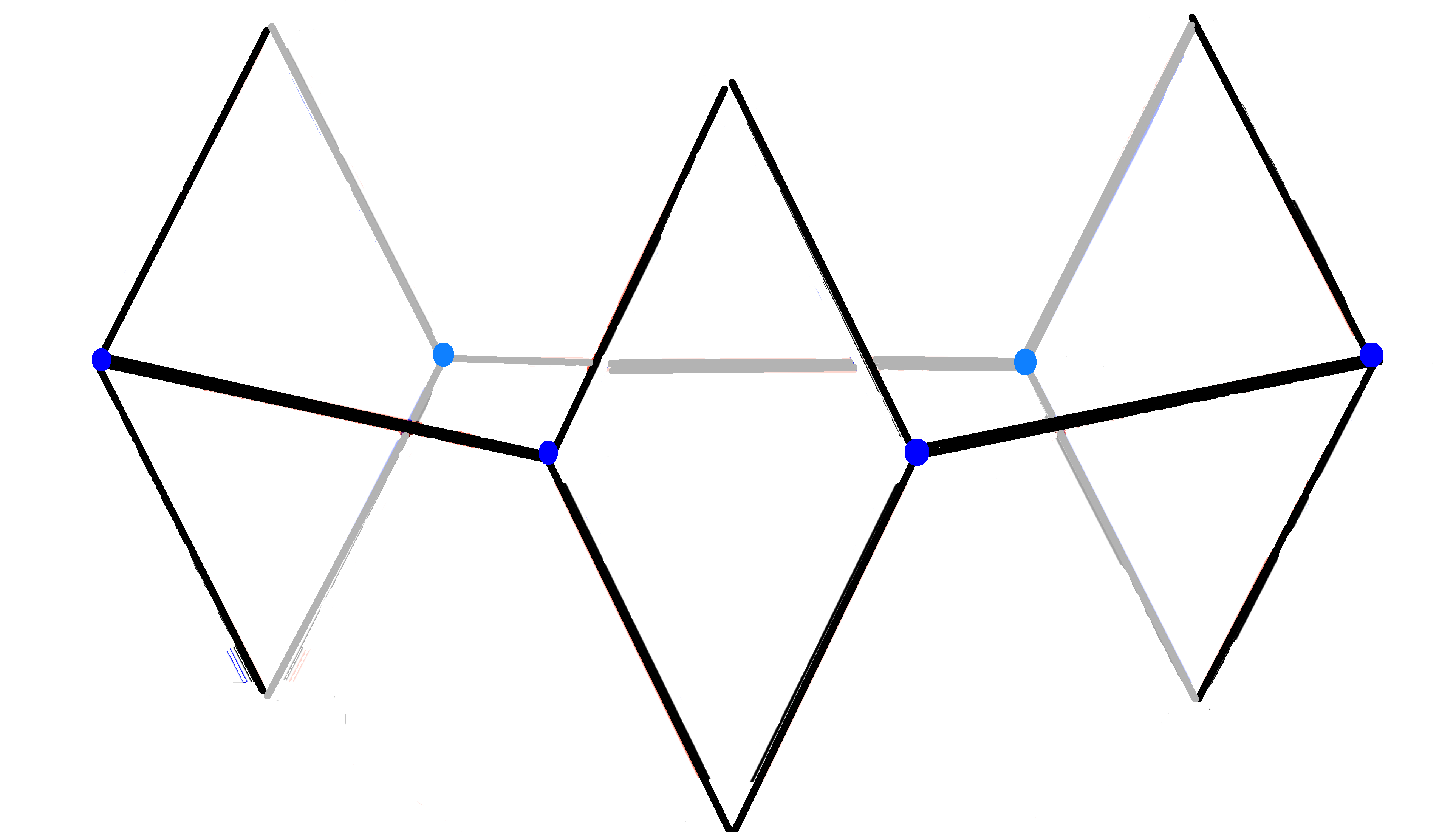}}
  \centering
   
\hspace{0.5cm}
\subfloat[$A_4$ necklace made with $A_1\times A_1$ quadrilateral beads.\label{fig:sub2}]{\includegraphics[width=.38\linewidth]{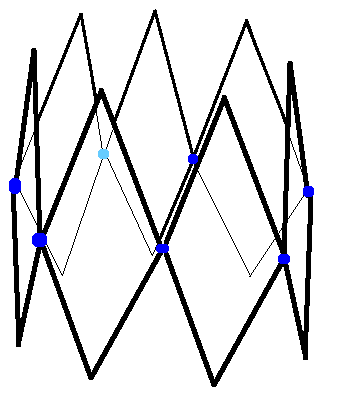}}
  \centering
  \caption{ $A_3$ and $A_4$ diamond necklaces living inside  their corresponding  Stasheff polytope. The blue vertices correspond to snake triangulations.}
\label{fig:test}
\end{figure}

 All this information will be relevant for us to define the set of multiplicatively independent \XX-coordinates below. In here we proceed looking further to a snake cluster/triangulation. All snake triangulations are equivalent by cyclic rotation of the polygon ${\bf P}_{n+3}$ for $n$ even and in the Stasheff polytope this can be seen by a diamond necklace with identical diamond beads of type $(A_1)^{\tfrac{n}{2}}$ formed when we freeze alternating diagonals of a snake triangulation. For $n$ odd, cyclic rotation defines two classes of snake triangulations that relate to one another by a dihedral symmetry. This can be seen  at the diamond necklace in the Stasheff polytope where for $n$ odd, there are a two distinct diamond beads $d_1$ and $d_2$ of type $(A_1)^{\tfrac{n-1}{2}}$ and $(A_1)^{\tfrac{n+1}{2}}$ respectively (figure \ref{fig:sub1}). $d_1$ and $d_2$ appear in pairs next to each other $\tfrac{n+3}{2}$ times in the necklace. 
 
\begin{figure}[ht] 
\begin{center} 
 \includegraphics[width=7cm]{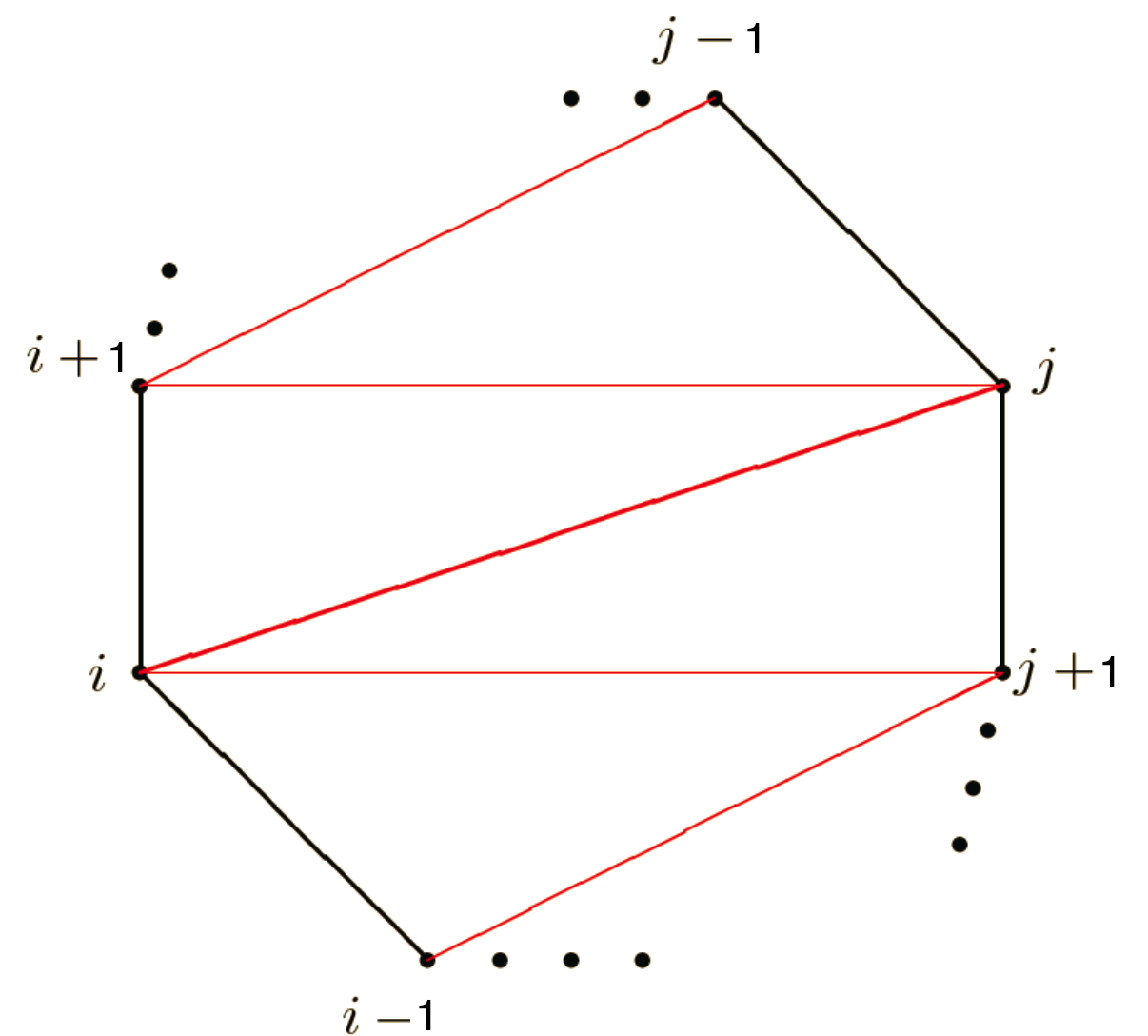}
\end{center} 
\caption{$\overline{ij}$ diagonal in a ``snake''  triangulation of an n-gon} 
\label{fig:n-snake} 
\end{figure}

 The relevant cluster algebras for us in the next section will be given by setting all cluster coefficients to 1. In this case, for any given cluster variable $\vev{ij}, (i+1<j\leq n+3)$, there is a snake triangulation shown in figure \ref{fig:n-snake}  that contain the diagonal $\overline{ij}$ that splits the quadrilateral $i(i+1)j(j+1)$ into two triangles. In this quadrilateral, the Pl\"ucker relation is given by
\begin{equation}\label{Plucker snake}
\vev{i,j}\vev{i+1,j+1}=1+\vev{i,j+1}\vev{i+1,j},
\end{equation}
 and the \XX-coordinate corresponding to the diagonal $\overline{ij}$ in this triangulation is 
 \begin{equation}\label{snake x-coord}
 x_{ij}= \frac{1}{\vev{i+1,j}\vev{i,j+1}},
 \end{equation}
 and in case $j=i+2$ both equations simplify a little further with $\vev{i+1,j}=1$. There are $\frac{n(n+3)}{2}$ of these snake \XX-coordinates (up to inversion) coming from all snake clusters, matching the number of cluster variables. In fact, from equation \ref{snake x-coord}, the snake \XX-coordinates can be seen as a  reparametrization of the cluster variables and therefore they are multiplicatively independent in the same way cluster variables are by general conditions of the Laurent phenomena in cluster variables.


\section{Kinematic space and its \texorpdfstring{$A_n$}{An} cluster structure}
\label{cluster kinematical}

The kinematic space $\mathcal{K}_m$ of a planar $m$-point amplitude of a massless particle is given by the  Mandelstam variables
\begin{equation}
s_{ij}= (p_i+p_j)^2.
\end{equation}
But only  $\frac{m(m-3)}{2}$ of them are independent for $m>3$ in four dimensional spacetime.

In \cite{Arkani-Hamed:2017mur}, a basis for $\mathcal{K}_m$ was defined by the set of planar kinematic variables:
\begin{equation}
X_{ij}= (p_i+p_{i+1}+\dots+p_{j-1})^2= (p_j+p_{j+1}+\dots+p_m+\dots+p_{i-1})^2= X_{ji},
\end{equation}
 where we used momentum conservation and $i<j<m$ and the indices are taken {\bf mod} m. Considering that $X_{i,i+1}=0$, there are  $\frac{m(m-3)}{2}$ planar kinematic variables that span $\mathcal{K}_m$.

In \cite{Arkani-Hamed:2017mur}, the scattering form $\Omega(\mathcal{A}_m)$ is a $(m-3)$-form, matching the number of propagators of a planar tree-level $m$-point diagrams, and it sums over all planar Feynman diagrams.

\begin{equation}
\Omega(\mathcal{A}_m)= \left(\sum_{\text{diagr.}}\frac{1}{\prod_{a=1}^{a=m-3}X_{i_a,j_a}}\right)d^{m-3}X
\end{equation}

Each planar diagram correspond to a triangulation of a polygon $\mathbf{P}_m$ and the respective set of $m-3$ planar variables $\{X_{i_a,j_a}\}$ has a correspondence to the diagonals $\{\overline{i_aj_a}\}$ in this triangulation. A sum over all such diagrams corresponds to a sum over the $A_{m-3}$ Stasheff polytope.

Many aspects of a $A_{m-3}$ cluster algebra already appears but it is not quite so. Mutation rules given by Pl\"ucker relations are (partially) missing, as we will see in a moment.

In order to reduce the space of integration in $\mathcal{K}_m$ to a section of $(m-3)$ dimensions $s_{ij}$ is set to be a negative constant for every non-adjacent indices $1\leq i<j\leq m-1$.

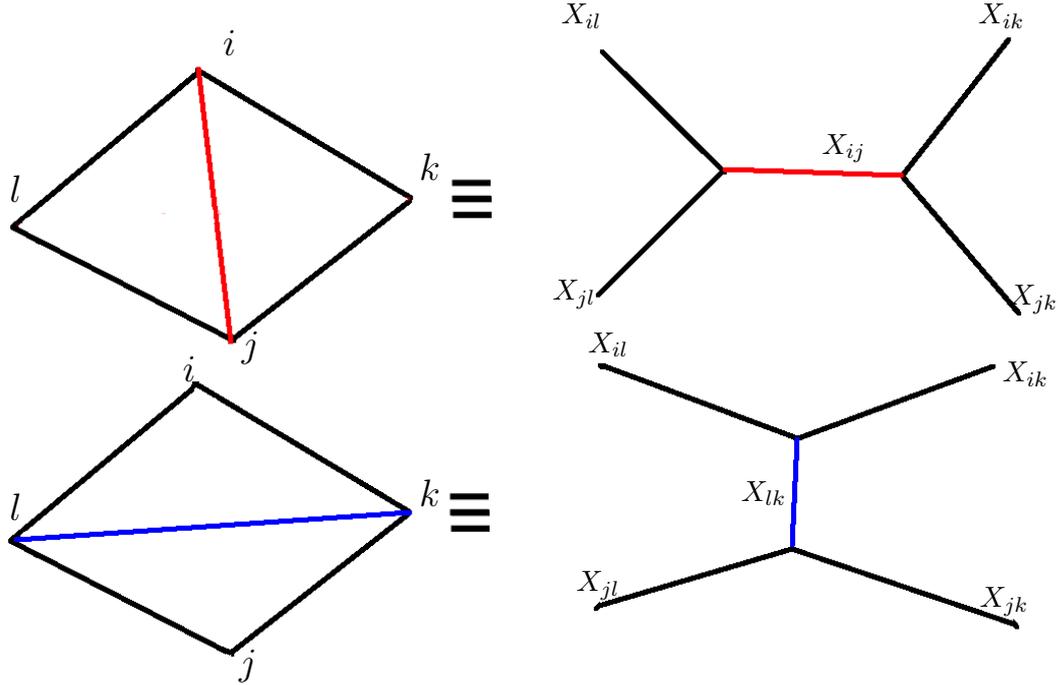
\begin{figure}
  \begin{picture}(200,250)
  \put(0,0){\includegraphics[width=13.5truecm]{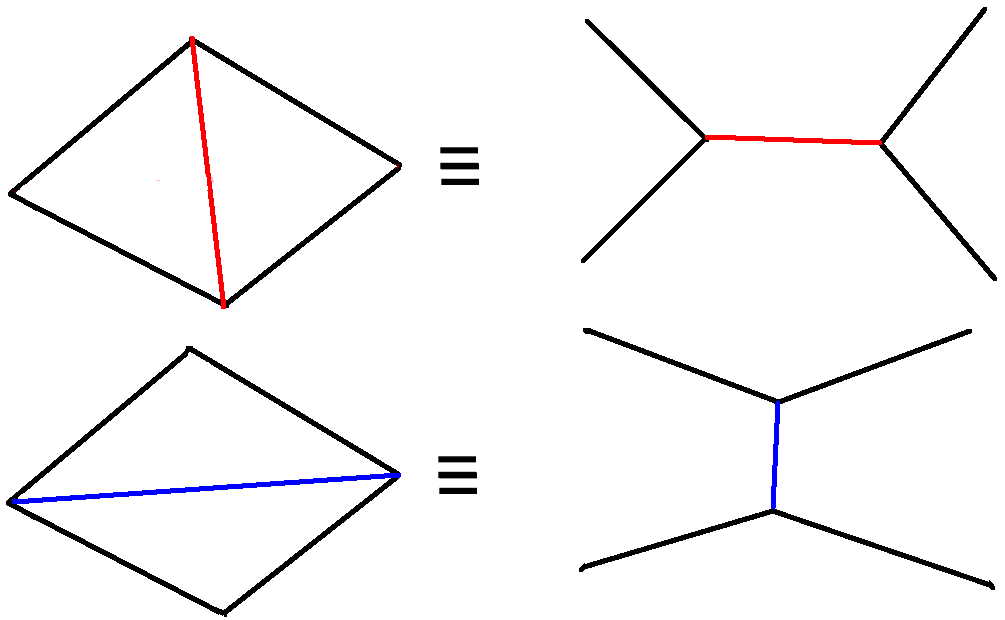}}
  \put(45,102){\makebox(50,15){\Large $i$}}
  \put(67,-10){\makebox(50,15){\Large$j$}}
  \put(-20,50){\makebox(50,15){\Large$l$}}
  \put(135,55){\makebox(50,15){\Large$k$}}
\put(358,100){\makebox(50,15){$X_{ik}$}}
\put(200,111){\makebox(50,15){ $X_{il}$}}
\put(198,20){\makebox(50,15){$X_{jl}$}}
\put(350,14){\makebox(50,15){$X_{jk}$}}
\put(260,55){\makebox(50,15){$X_{lk}$}}
\put(60,225){\makebox(50,15){\Large $i$}}
  \put(68,110){\makebox(50,15){\Large$j$}}
  \put(-20,170){\makebox(50,15){\Large$l$}}
  \put(135,178){\makebox(50,15){\Large$k$}}
\put(349,235){\makebox(50,15){$X_{ik}$}}
\put(190,235){\makebox(50,15){ $X_{il}$}}
\put(189,130){\makebox(50,15){$X_{jl}$}}
\put(361,128){\makebox(50,15){$X_{jk}$}}
\put(290,185){\makebox(50,15){$X_{ij}$}}
  \end{picture}
  \caption{Mutation of $X_{ij}$ to $X_{lk}$, resembling tropical cluster mutation.}
  \label{ikjl mutation}
\end{figure}

Mutation relations  (figure \ref{ikjl mutation}) associated to the quadrilateral $ikjl,\;(1\leq i<j<k<l<m)$ are
 
 \begin{equation}
 X_{ij}+X_{kl}= X_{il}+X_{kj}+ \textnormal{constants}.
 \end{equation}
 This is not a Pl\"ucker relation (eq. \ref{Plucker}). The product is replaced by a sum and this can be seen as a tropical version of the exchange relation, similar to the tropical exchange relation in \cite{Fomin:hy} between the exponent of the denominators of the cluster variables when written in terms of cluster variables of a initial cluster. But a term is missing in the expression above and therefore, 
the $A_{m-3}$ cluster algebra is not completely realized in a tropical form.
But we can see that in all snake triangulations, all quadrilaterals made of adjacent triangles (figure \ref{fig:n-snake}) has its Pl\"ucker relations (eq. \ref{Plucker}) with one term replaced by 1 (\ref{Plucker snake}). Furthermore, all exchange relations among clusters in the diamond necklace sector of the Stasheff polytope are like eq. \ref{Plucker snake}. This can be seen as a residual cluster algebra.

The realization of such residual cluster algebra starts by looking in a quadrilateral $\overline{i(i+1)j(j+1)}$ (figure \ref{fig:n-snake}), $i+1<j$. We have the following relations:\begin{align}
X_{ij}+X_{i+1,j+1} &= X_{i+1,j}+X_{i,i+1} +cte.\label{tropical cluster}\\
\vev{ij}\vev{i+1,j+1}&= \vev{i+1,j}\vev{i,j+1}+1 \label{Plucker snake2}
\end{align}

So in a kinematic limit where $X$'s are much bigger than 1 and the constants $s_{ij}$ ($i$ and $j$ non-adjacent) we see that eq. \ref{tropical cluster} is a realization of the exchange relation \ref{Plucker snake2} by defining cluster variables as 
\begin{equation}
\vev{ij}= \exp(X_{ij}),\quad\quad (\vev{ij}>>1).
\end{equation}
This kinematic limit portraits a high energy one, possibly off-shell and not always applicable in a tree level computation. What is important is to know that there is present in $\mathcal{K}_m$ the structure of a residual cluster algebra that manifests its presence in the amplitude.
 


\section{\texorpdfstring{$\mathcal{K}_n$}{Kn} vs. \texorpdfstring{$\Conf_n(\PP^3)$}{Conf(\PP3)}}\label{Kn vs Confn}
$\mathcal{K}_n$ and $\Conf_n(\PP^3)$ are different parametrization spaces of the  kinematic data of $n$ external particles written in terms of momentum and momentum twistors, respectively. Therefore $\mathcal{K}_n$ suits for theories without supersymmetry while $\Conf_n(\PP^3)$ takes advantage of the conformal symmetries of  $\mathcal{N}=4$ SYM. Furthermore, while $\mathcal{K}_n$ has a  residual cluster algebra  of $A_{n-3}$ type whose cluster exchange (mutation) relations survive only along the diamond necklace inside the $A_{n-3}$ Stasheff polytope, $\Conf_n(\PP^3)$ has a larger and fully preserved cluster algebra $\Gr(4,n)$ that is of type $A_3$ for $n=6$, $E_6$ for $n=7$ and of infinite type for $n\geq 8$.

What they have in common is that the $\Gr(4,n)$ cluster variables and the $A_{n-3}$ tropical cluster variables are used to build the same physical objects, the  cross-ratios. Cross-ratios are conformally invariant and this property makes them a good choice for entry kinematic data in the polylogarithm expressions of the remainder functions  in $\mathcal{N}=4$ SYM amplitudes. 

Disregarding Gram determinant and on-shell conditions, there are $\frac{n(n-3)}{2}$ independent\footnote{$n$ of them are zero at $j=i+2$ when we impose on-shell condition.} standard cross-ratios written in terms of planar kinematic variables (our tropical \A-coordinates) \cite{Drummond:2007au,Anastasiou:2009kna}
\begin{equation}
u_{ij}=\frac{X_{i,j+1}X_{i+1,j}}{X_{ij}X_{i+1,j+1}}
\end{equation}
with $1<i+1<j\leq n$ and cyclicity of the indices applied when necessary. Written in terms of Pl\"ucker coordinates of $\Gr(4,n)$
\begin{equation}\label{crossratio}
u_{ij}=\frac{\vev{i,i+1,j+1,j+2}\vev{i+1,i+2,j,j+1}}{\vev{i,i+1,j,j+1}\vev{i+1,i+2,j+1,j+2}}=\frac{\vev{i,j+2,I}\vev{i+2,j,I}}{\vev{i,j,I}\vev{i+2,j+2,I}}
\end{equation}
where we isolated and freezed the commom momentum-twistors indices  in $I=\{i+1,j+1\}$. In this way only $n-2$ external points are considered and $i$ and $i+2$ become adjacent and similarly for $j$ and $j+2$. We can see using Pl\"ucker relations \ref{Plucker} that the inverse of expression (\ref{crossratio}) becomes  $u_{ij}^{-1}= x_{ij}^{-1}+1$  where $x_{ij}$ is a snake \XX-coordinate. 

The common thread of the cross ratio written in terms of $A_{n-3}$ tropical \A-coordinates in $\mathcal{K}_n$ and $\Gr(4,n)$ \XX-coordinates in $\Conf_n(\PP^3)$ is that these coordinates belong to  snake clusters in $A_{n-3}$ residual cluster algebra of $\mathcal{K}_n$ and in $A_{n-5}\subset \Gr(4,n)$ cluster sub-algebras\footnote{These are distinct cluster sub-algebras, according to the suppressed pair of indices $I$.} of $\Conf_n(\PP^3)$, respectively. At $n=6$ and 7, the isomorphisms $\Gr(4,6)\equiv \Gr(2,6)$ and $\Gr(4,7)\equiv \Gr(3,7)$ enhance the cluster sub-algebras $A_{n-5}$ in $\Gr(4,n)$ to $A_3$. 

\section{Conclusion}\label{conclusion}

We found a residual $A_{n-3}$ cluster algebra in the kinematic space $\mathcal{K}_n$, $n>4$ with planar kinematic variables $X_{ij}$ playing the roles of tropical cluster variables. It is residual because the exchange relations in its tropical form are in accordance with a cluster algebra of type $A_{n-3}$ only in a subset of all clusters that form what we call a diamond necklace inside the $A_{n-3}$ Stasheff polytope, or generalized associahedron. This necklace is formed by $n$ vertices corresponding to snake clusters and hypercubes (diamonds) that are powers of $A_1$ cluster sub-algebras that appear when we freeze cluster variables corresponding to alternating diagonals in a snake triangulation.  

Nevertheless, there are exchange relations for all $X_{ij}$ (eq. \ref{tropical cluster}) represented geometrically by the exchange of diagonals in a quadrilateral in the same way as in a $A_{n-3}$ cluster algebra and this exchange relations are the edges that builds an $A_{n-3}$ associahedron. Due to these exchange relations outside the diamond necklace, Laurent phenomena fails and these residual cluster variables (or its tropical ones) cannot be related to the root system of $A_{n-3}$. Therefore, the original \cite{Fomin:hy} tropical exchange relation of the exponent of the denominators of a cluster variable $\vev{ij}$ will make no sense at all.

In \cite{Torres:2013vba}, we already noticed that in the reduced special 2D kinematic configuration space that the remainder function of a two loop $\mathcal{N}=4$ SYM MHV amplitude should depend only of \XX-coordinates of snake clusters of type $A$ cluster algebras.  These snake \XX-coordinates are  squared geometric mean of the adjacent cluster variables (\A-coodinates) (eq. \ref{snake x-coord}) and they are multiplicatively independent if the cluster variables are. The multiplicative independence  could be checked by hand for $A_3$ cluster algebra with coeficients set to 1 by writing the cluster variables $x_{ij}$ as Laurent Polynomials 
\begin{equation} 
x_{ij}=\frac{P_{ij}(a_1,a_2,a_3)}{a_1^la_2^ma_3^n}  
\end{equation}
in three cluster variables $a_1, a_2, a_3$ of a initial cluster. The denominator of $A_n$ cluster variables are always a monomial in powers of the initial cluster variables that  pair with the root system of $A_n$ Lie Algebra \cite{Fomin:hy,Fomin:ky}. We believe that in the general case due to Laurent phenomena and with general values for the coefficients, the snake cluster variables and snake \XX-coordinates of  $A_n$ cluster algebras will be multiplicative independent.

Coincidentally, as we showed in section \ref{Kn vs Confn}, even before the language of cluster algebras started being used in scattering amplitudes,  the literature shows that the canonical choice of cross-ratios are direct expressions of the snake \XX-coordinates of type $A$ cluster sub-algebras in $\Conf_n$ and tropical coordinates of snake clusters in $\mathcal{K}_n$.  

Snake clusters coordinates of a $A$ cluster sub-algebra seam to be the safest choice of coordinates, where some properties of the cluster algebra are still preserved when physical circumstances (loop expansion, helicity, lack of supersymmetry) break it to a residual cluster algebra in its diamond necklace.  

\acknowledgments
The author thanks IMPA for the hospitality while this article was written, and specially Prof. Hossein Movasati for the kind invitation to visit him at IMPA.






\bibliographystyle{unsrtnat}
\bibliography{ClusterAlgebras}



\end{document}